\documentclass[aps,prl,twocolumn,superscriptaddress,showpacs]{revtex4-1} 
\usepackage{graphicx}
\usepackage{amsmath}
\usepackage{epstopdf}
\usepackage{hyperref}
\begin{document}
\newcommand{\<}{\langle}
\renewcommand{\>}{\rangle}
\newcommand{\beq}{\begin{equation}}
\newcommand{\eeq}{\end{equation}}
\newcommand{\kc}{K_{\text{c}}}
\newcommand{\lt}{L_\tau}
\newcommand{\asp}{\alpha_{\tau}}
\newcommand{\br}{{\bf r}}
\newcommand{\bk}{{\bf k}}
\newcommand{\bJ}{{\bf J}}

\title{Quantum Critical Dynamics Simulation of Dirty Boson Systems}

\author{Hannes Meier} 
\author{Mats Wallin} 
\affiliation{Department of Theoretical Physics, KTH Royal Institute of
  Technology, SE-106 91 Stockholm, Sweden}

\date{\today}

\begin{abstract}
  Recently the scaling result $z=d$ for the dynamic critical exponent
  at the Bose glass to superfluid quantum phase transition has been
  questioned both on theoretical and numerical grounds.  This
  motivates a careful evaluation of the critical exponents in order to
  determine the actual value of $z$.  We study a model of quantum
  bosons at $T=0$ with disorder in 2D using highly effective worm
  Monte Carlo simulations.
  Our data analysis is based on a finite size scaling approach to
  determine the scaling of the quantum correlation time from
  simulation data for boson world lines.  The resulting critical
  exponents are $z=1.8 \pm 0.05, \nu=1.15 \pm 0.03,$ and $\eta=-0.3
  \pm 0.1$, hence suggesting that $z=2$ is not satisfied.
\end{abstract}


\maketitle

Quantum phase transitions (QPT) occur at zero temperature and produce
new and important physics compared to ``classical'' phase transitions
at finite temperature \cite{Sondhi:1997p173,Sachdevbook}.  In
particular, presence of quenched disorder leads to new universality
classes without direct classical counterparts, where lack of
space-time symmetry can lead to nontrivial scaling properties.  Such
phenomena are of great current interest and present considerable
theoretical and experimental challenges
\cite{Sondhi:1997p173,Sachdevbook}.

A prototype QPT with disorder is the 2D boson superfluid to insulator
transition in the presence of random substrate disorder.  The disorder
localized insulating phase is a gapless phase called the Bose glass.
This transition is relevant for experiments on ultrathin granular
superconducting films, Josephson junction arrays, superfluid helium
films, and cold bosons in optical lattices with disorder
\cite{Fisher:1989p868,White:2009p4815}.  A remarkable result of the
theory is the relation $z=d$, where $d$ is the number of spatial
dimensions \cite{Fisher:1989p868}.  
This scaling result was believed to be exact, but has been questioned
recently both analytically and numerically
\cite{Weichman:2007p1191,Weichman:2008p1051,Priyadarshee:2006p740}.
The result $z=d$ is derived by requiring the contribution to the
compressibility $\kappa$ from the singular part of the free energy to
be a nonsingular function across the transition.  However, if $\kappa$
instead comes from the analytic part of the free energy no restriction
on $z$ follows and the relation $z=d$ does not have to hold
\cite{Weichman:2007p1191}.
In 1D $z=1$ is fulfilled \cite{Fisher:1989p868},
but Ref.\ \cite{Weichman:2008p1051} finds that this is 
unrelated to the mechanism that keeps the compressibility finite
through the Bose glass-superfluid transition.

The task of determining the quantum dynamical exponent at the
disordered boson QPT to test the validity of the relation $z=2$ in 2D
has been studied previously.  An often used approach has been to
assume the value $z=2$ and then test if scaling can be obtained by
fitting other parameters to numerical data.  This approach produces
seemingly good scaling results for the system sizes tested
\cite{Sorensen:1992p94,Wallin:1994p924,Alet:2003p741,Lin:1107.3107v1},
but does not rule out that a calculation without a priori assumptions
might give a different result.  
A recent simulation study reports $z\approx 1.4$
\cite{Priyadarshee:2006p740}, but this result might be affected by the
limited disorder averaging used \cite{Lin:1107.3107v1}.
Renormalization approaches have also been used to determine $z$
\cite{Herbut:2000p4126}, but have not yet settled
\cite{Weichman:2007p1191}.  Thus the validity of the result $z=d$ is
unclear and further tests are required.

In this paper we perform large scale Monte Carlo (MC) simulations to
determine $z$ and other critical exponents at the Bose glass
transition of the dirty boson model in $d=2$ dimensions.  We extend
previous simulation results in several ways.  We use extensive
disorder averaging, and larger system sizes than in most previous
studies, which turns out to be crucial.  A highly effective worm
algorithm is used that permits efficient averaging over configurations
with different boson winding numbers \cite{Prokofev:2001p1358}.  
In order to locate the QCP and study dynamical scaling, a suitable
function of the winding number is constructed that has a maximum value
when the system size in the time direction is proportional to the
correlation time.  Finite size scaling of the maximum gives a direct
route to calculating $z$ and other critical exponents, without any a
priori assumptions on $z$.  The results display significant
corrections to scaling for small system sizes that complicates
determination of the exponents.  Our estimate, $z=1.8 \pm 0.05$,
suggests that the dynamic exponent is smaller than given by the
relation $z=d$ for $d=2$.  

{\em Dirty boson model} -- The imaginary-time path-integral
representation of the 2D Boson Hubbard model with nearest neighbor
hopping, on-site charging energy, and a disordered chemical potential
can be mapped to a link-current model convenient for simulation
\cite{Wallin:1994p924}.  The link-current model assumes only phase
fluctuations of the order parameter and neglects amplitude
fluctuations, has isotropic space-time couplings, and uses the Villain
form of the potential \cite{Wallin:1994p924}.  Such details are not
expected to alter the universality class of the QPT.  The Hamiltonian
of the link-current model is
\beq 
H =\frac{1}{K} \left( \sum_{i,\delta}\frac{1}{2} (J^\delta_{i})^2- \sum_{i}
  (\mu+v_{\br}) J_i^{\tau} \right)
\label{eq:Hamiltonian}
\eeq 
Here $i=(\br,\tau)$ denotes the sites of a $(2+1)$-dimensional simple
cubic spacetime lattice of size $L\times L \times L_{\tau}$ with
periodic boundary conditions in space and time directions, and
$\delta=x,y,\tau$ denotes the coordinate directions.  The integer link
variables $J^\delta_{i}$ represent boson current variables on the
links extending from the site $i$ in the $\delta$-direction.  The
variables are subject to the divergence-free constraint $\nabla \cdot
\bJ = 0$, which means that the worldlines have no open ends.  $K$ is a
coupling constant.  Disorder is modeled as a quenched on-site
potential which is random in space but constant in the time direction,
with a uniform distribution in $|v_{\br}|<1/2$.  The chemical
potential is here fixed to $\mu=1/2$ which means half filling of
bosons on average.  The transition of this model represents the
generic universality class of the disorder driven boson localization
QPT.

Next we introduce the two main quantities of interest in our
simulations.  The mean square winding number is defined as
\beq
W_{\delta}^2 = \left[ \left< \left( 
\frac{1}{L_\delta}\sum_i J^{\delta}_i \right)^2 \right> \right]
\label{eq:winding}  
\eeq
The bracket $\< \cdots \>$ indicates average with respect to
$J$-current configurations, and $[ \cdots ]$ indicates the quenched
disorder average.  The spatial mean square winding number measures
fluctuations in the number of times the worldlines wind across the
sample, and is proportional to the superfluid density
\cite{PhysRevB.36.8343}.  It can thus be used to detect the boson
superfluid to insulator QCP.  The temporal winding number fluctuations
(including subtraction of the average boson number) correspond to the
boson compressibility $\kappa$ \cite{Wallin:1994p924}.  The gapless
nature of the Bose glass produces a smooth nonzero compressibility
across the transition \cite{Fisher:1989p868}.  From now on we will
only consider spatial winding number fluctuations, and form
$W^2=(W_x^2+W_y^2)/2$.  The Greens function
$G(\br-\br',\tau-\tau')=[\langle
e^{i(\theta_{\br,\tau}-\theta_{\br',\tau'})}\rangle]$ can be used to
define the uniform order parameter susceptibility
$\chi=G(\bk=0,\omega=0)$ \cite{Wallin:1994p924,Fisher:1989p868}.

\begin{figure}
\includegraphics[width= 0.9\columnwidth]{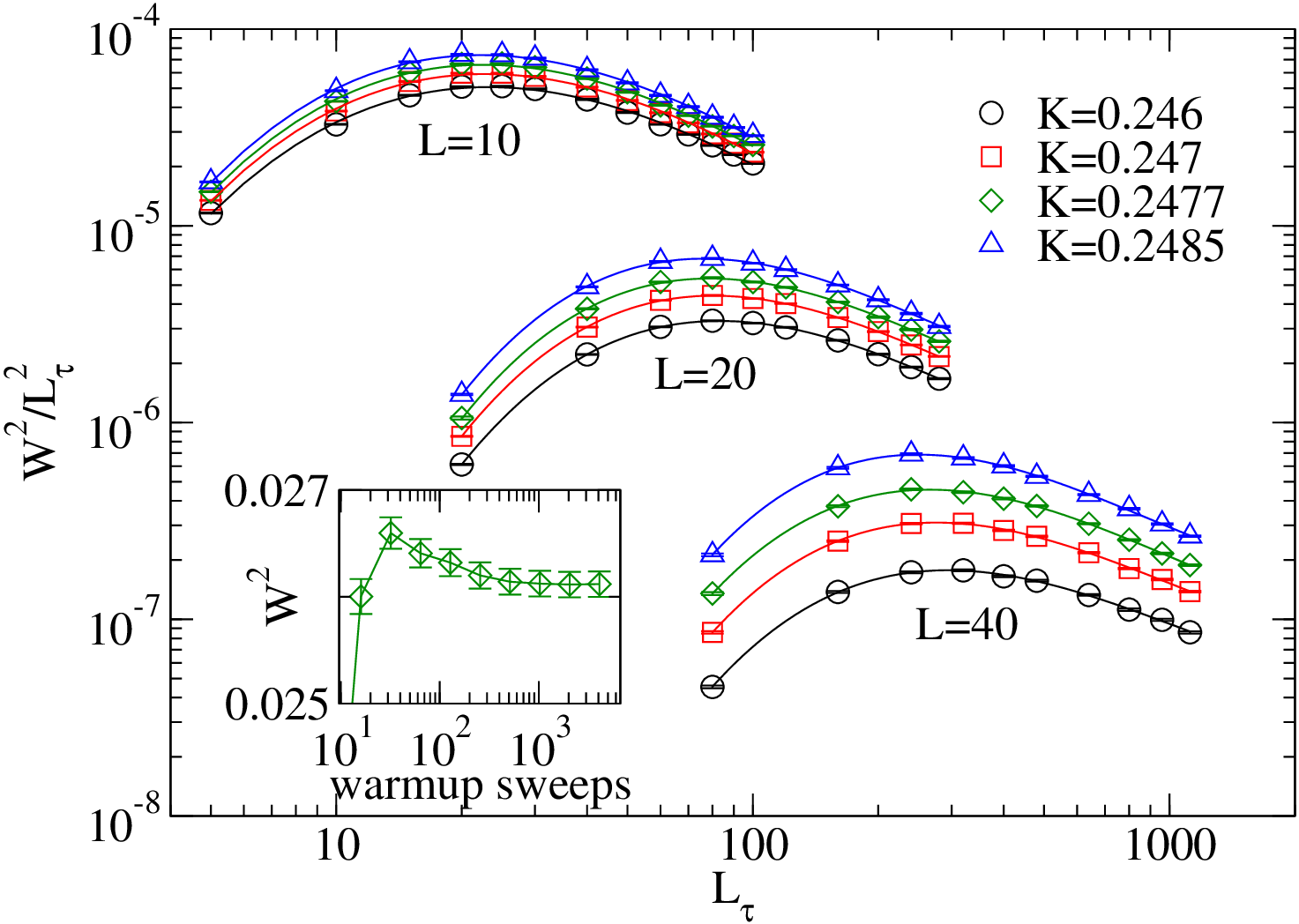}
\caption{Selection of Monte Carlo results for the winding number
  fluctuation divided by $L_{\tau}^{2}$ as a function of $\lt$.  Solid
  curves are polynomial fits to the data curves, from which the 
  locations $\lt^*$ and sizes $(W^2/\lt^{2})^*$ of the maxima can be
  determined.  Inset: Equilibration test for $L=40, \lt=240,
  K=0.2477$.}
\label{fig1}
\end{figure}

{\em Monte Carlo simulations} -- Our MC simulations use the classical
lattice worm algorithm \cite{Prokofev:2001p1358}.  For each disorder
realization the simulation was started in the $J$-current
configuration that minimizes $H$ in Eq.\ (\ref{eq:Hamiltonian}).  The
simulations used more than 1500 MC sweeps to reach equilibrium,
followed by equally many sweeps to collect data for the averages.
Here a MC sweep is defined as $3L^2L_\tau$ link variable update
attempts.  Measurements are taken every time the worm closes.  The
winding number is given by the number of times the world lines wrap
around the sample, and the susceptibility is the average number of
update attempts per closed loop configuration \cite{Alet:2003p741}.
We tested for equilibration by monitoring disorder averages of the
winding number fluctuations and of the susceptibility calculated using
different numbers of warmup sweeps. An example is shown in the inset
in Fig.\ \ref{fig1}.  The results become independent of the initial
configuration after about 500 warmup sweeps.  The quenched disorder
averaging used between $10^4-10^5$ samples of the random potential,
where more disorder averaging was used around the critical point.
Statistical error bars on the data points were estimated by
fluctuations in the disorder averages.

{\em Finite-size scaling methods} -- The basic scaling assumption is
that the correlation length and time diverge at the transition as $\xi
\sim |k|^{-\nu}$ and $\tau \sim \xi^z$, where $k=(K-\kc)/\kc$, $\kc$
is the critical coupling, $\nu$ is the correlation length exponent,
and $z$ is the dynamic exponent.  The winding number fluctuation is
dimensionless and therefore scale invariant at the transition.  We
assume the following finite size scaling (FSS) ansatz for the winding
number fluctuation
\beq
W^2 (K,L,\lt) = \tilde{W}^2 (L^{1/\nu}k,\asp)
\label{eq:wscaling}
\eeq
and for the susceptibility
\beq
\chi (K,L,\lt) = L^{2-\eta} \tilde{\chi} (L^{1/\nu}k,\asp)
\label{eq:chiscaling}
\eeq 
where $\tilde{W}^2$ and $\tilde{\chi}$ are scaling functions, and
$\asp = \lt/L^z$ is the aspect ratio.  The aim is to estimate the
critical exponents $z, \nu, \eta$ and scaling functions by fitting
these expressions to numerical MC data for finite $L,\lt$.

FSS analysis greatly simplifies if the scaling functions can be
reduced to functions of only one variable by taking the other variable
to be constant.  Taking the first variable $L^{1/\nu}k$ to be constant
means keeping $K=\kc$, which is a priori unknown, while keeping the
second variable constant requires knowledge of $z$.  Most previous
studies have therefore assumed the value $z=2$ and selected system
sizes for simulations given by $\lt = const \times L^2$.  Clearly this
approach is not available if the value of $z$ is unknown.

The idea is now to, without assuming knowledge of $\kc$ and $z$,
construct a characteristic scale $\lt^*$ for each given $K,L$, which
scales as $\lt^* \sim \tau \sim L^z$ for $K=\kc$, where $\tau$ is the
correlation time.  The winding number fluctuation is a monotonically
increasing function of $\lt$ for fixed $K,L$.  For $\lt \gg \tau$ the
worldline fluctuations separated by times greater than the correlation
time $\tau$ decorrelate, and then the winding number fluctuation must
increase linearly with $\lt$.  Thus the quantity $W^2/\lt$ approaches
a constant value for $\lt \gg \tau$.  Dividing once more gives
$W^2/\lt^2$, which has a maximum at a characteristic $\lt^*$, and goes
to zero for $\lt \gg \tau$, where the star indicates the value at the
maximum.  We will find these maxima very useful in the scaling
analysis \cite{Vestergren:2004p3124}.  A convenient scaling form is
produced by replacing $\lt$ in $W^2/\lt^2$ by $\asp=\lt/L^z$.  We thus
introduce
\beq
\Phi(K,L,\lt) \equiv \frac{W^2}{\asp^2} = \tilde{\Phi} (L^{1/\nu} k, \asp)
\label{eq:phiscaling}
\eeq
This FSS relation is used below to estimate the critical coupling
$\kc$ and the exponents $z, \nu$.   
We verified that our approach reproduces known exponents for pure models.

\begin{figure}
\includegraphics[width= 0.9\columnwidth]{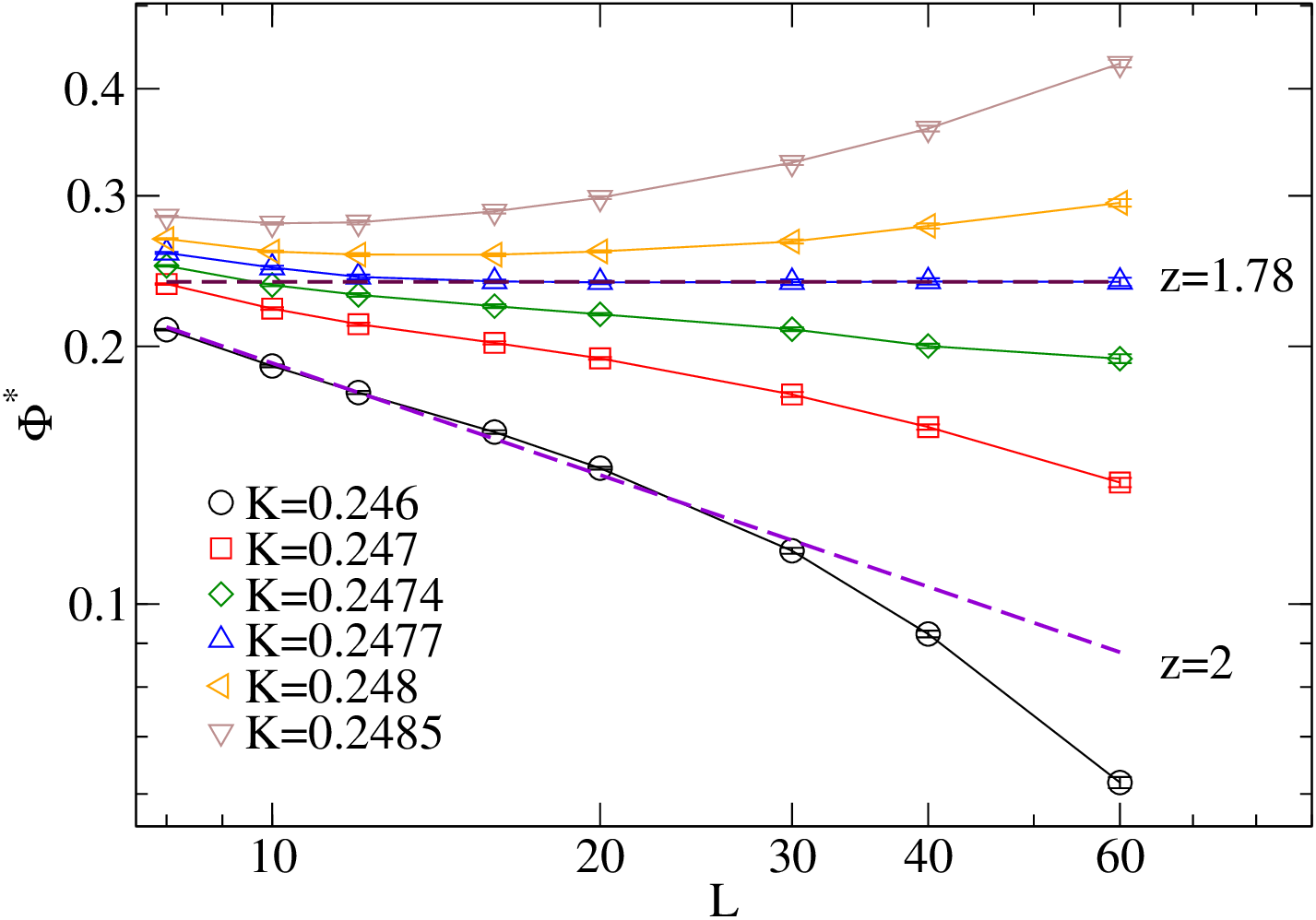}
\caption{Maximum values $\Phi^*=(W^2/\asp^2)^*$ vs.\ system size $L$
  for different couplings $K$.  For $z=1.78$ the data obeys
  $\Phi^*=const$ at $K=K_c$ for $L \ge 16$, indicated by the
  horizontal dashed line, which estimates $K_c=0.2477$.  The lower
  dashed line corresponds to $z=2$, which approximately describes the
  data for small sizes $L<16$ at $K=0.246$.  }
\label{fig2}
\end{figure}

{\em Results} -- First we locate the critical coupling $\kc$ and the
dynamic exponent $z$ by FSS analysis of MC data for the winding
number.  Figure \ref{fig1} shows examples of maxima of the quantity
$W^2/\lt^2$.  The amplitude $(W^2/\lt^2)^*$ and location $\lt^*$ of
the maxima can be straightforwardly computed by polynomial fits to the
MC data curves.  Better accuracy is obtained in the estimates for
$(W^2/\lt^2)^*$ than for $\lt^*$.  The maximum values scale as
$(W^2/\lt^2)^* \sim L^{-2z}$ at $K=K_c$.  However it is more
convenient to plot the quantity $\Phi^* = (W^2/\asp^2)^*$ of Eq.\
(\ref{eq:phiscaling}), and look for the scaling $\Phi^*\sim L^0$ at
$K=K_c$, which is shown in a log-log plot in Fig.\ \ref{fig2}.  In the
figure the of value $z$ enters through $\asp^* = \lt^*/L^z$, and has
been adjusted to make $\Phi^*=const$ at $K=K_c$ for system sizes
$L>16$, marked with the horizontal dashed line.  This produces the
estimates $z \approx 1.8$ and $K_c \approx 0.2477$.  For $K \ne \kc$
the data curves clearly splay out, away from a critical power law.
For $L<16$ deviation from power law behavior is obtained, which
indicates the presence of corrections to scaling in these data points.
In Fig.\ \ref{fig2} we also note that the choice $z=2$ gives an
approximate description of the data for $K=0.246$ for small system
sizes, $L<16$, which is indicated by the Ê lower dashed line, in
agreement with Ref.\ \cite{Alet:2003p741}.  As a consistency test, a
similar analysis was done for the location $\lt^*$ of the maxima using
the relation $\lt^* \sim L^z$ at $K=K_c$, leading to similar results.

\begin{figure}
\includegraphics[width=0.9\columnwidth]{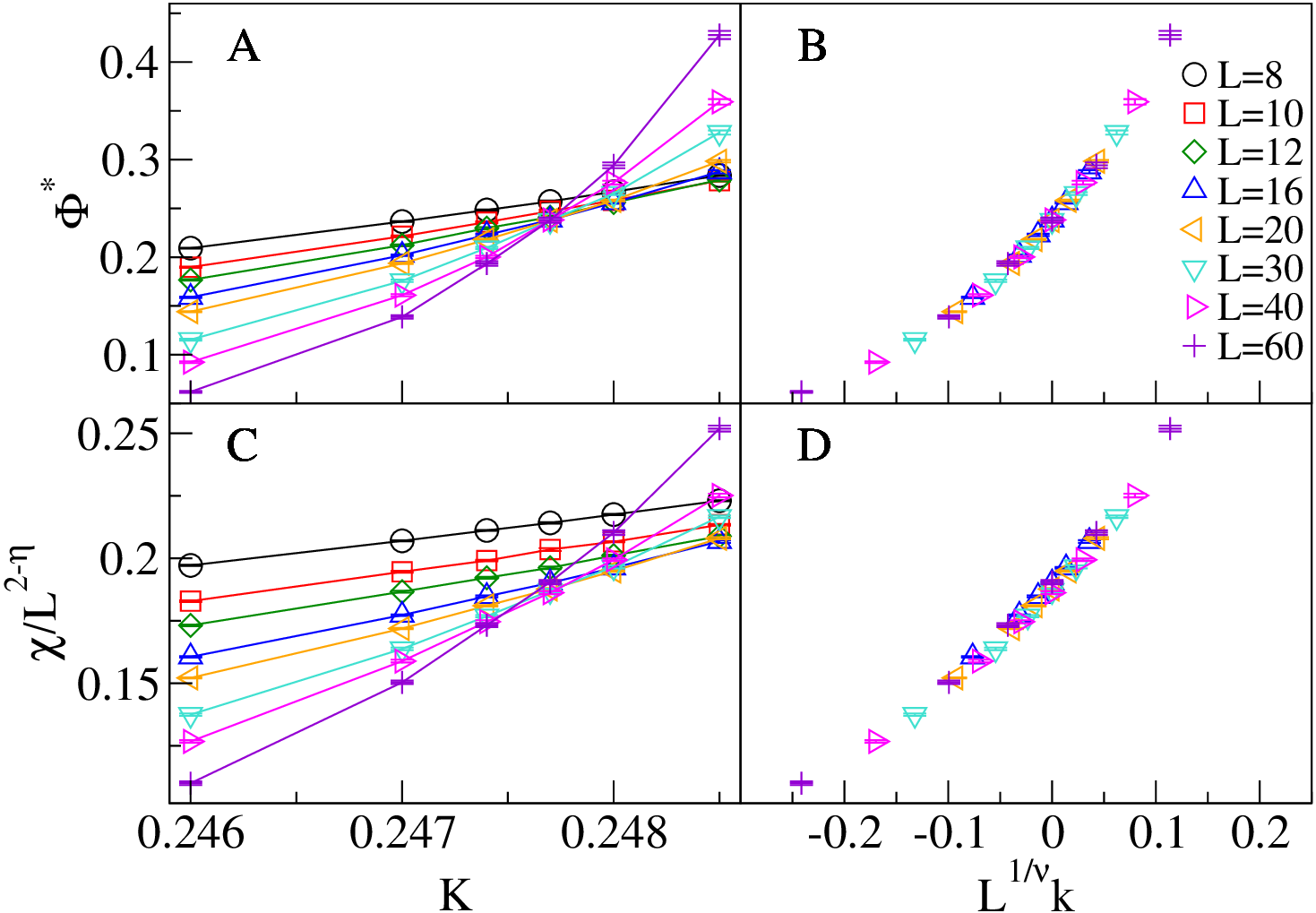}
\caption{A: Intersection plot for the scaled winding number function
  maxima $\Phi^*$, showing an intersection at $K_c=0.2477$.  B: FSS
  data collapse of the data in A obtained for $\nu=1.15, L \ge 16$.
  C: Intersection plot for the scaled susceptibility $\chi/L^{2-\eta}$
  evaluated at $\asp=0.35$.  The data curves for large sizes roughly
  intersect at $K=0.2477$ for $\eta=-0.29$, but with much larger
  corrections visible for small system sizes than for $\Phi^*$.  D:
  FSS data collapse of $\chi/L^{2-\eta}$ for $\nu=1.15, L \ge 16$.}
\label{fig3}
\end{figure}

Figure \ref{fig3} A shows the maxima of the function $\Phi^*$ of Eq.\
(\ref{eq:phiscaling}), with $\alpha=\lt/L^z$ for $z\approx 1.8$.  The
data curves for $L>16$ intersect at $K_c$, but for smaller sizes
scaling deviations are present, and these will be further discussed
below.  The correlation length exponent is readily estimated by
computing the derivatives $\partial \Phi^*/\partial K |_{K=K_c} \sim
L^{1/\nu}$, and a polynomial fit to the MC data gives $\nu \approx
1.15$.  The FSS data collapse produced by using this value for $\nu$
is shown in Fig.\ \ref{fig3} B for $L>16$.

To estimate the correlation function exponent $\eta$ we use the
susceptibility $\chi$ given by Eq.\ (\ref{eq:chiscaling}).
We fix the aspect ratio to $\asp = 0.35$ which correspond to the 
maxima $\Phi^*$ at criticality.  The value of
$\chi$ at this aspect ratio was determined by a polynomial fit 
to nearby MC data.  From $\chi \sim L^{2-\eta}$ we
estimate $\eta \approx -0.3$ for $L \ge 16$.  Figure \ref{fig3} C
shows a corresponding intersection plot for the quantity
$\chi/L^{2-\eta}$, which becomes size independent at $K=K_c$ according
to Eq.\ (\ref{eq:chiscaling}).  A FSS collapse assuming $\nu=1.15$ is
shown in Fig.\ \ref{fig3} D.  Note that the deviations from scaling
for small system sizes in Fig.\ \ref{fig3} C are substantial, and
hence the uncertainty in the estimate of $\eta$ is considerable.  The
scatter among the intersection points can be reduced by assuming a
scaling correction proportional to $L^{-\omega}$ with $\omega \approx
1$, but the accuracy of the data is insufficient for detailed
estimates.

\begin{figure}
\includegraphics[width=\columnwidth]{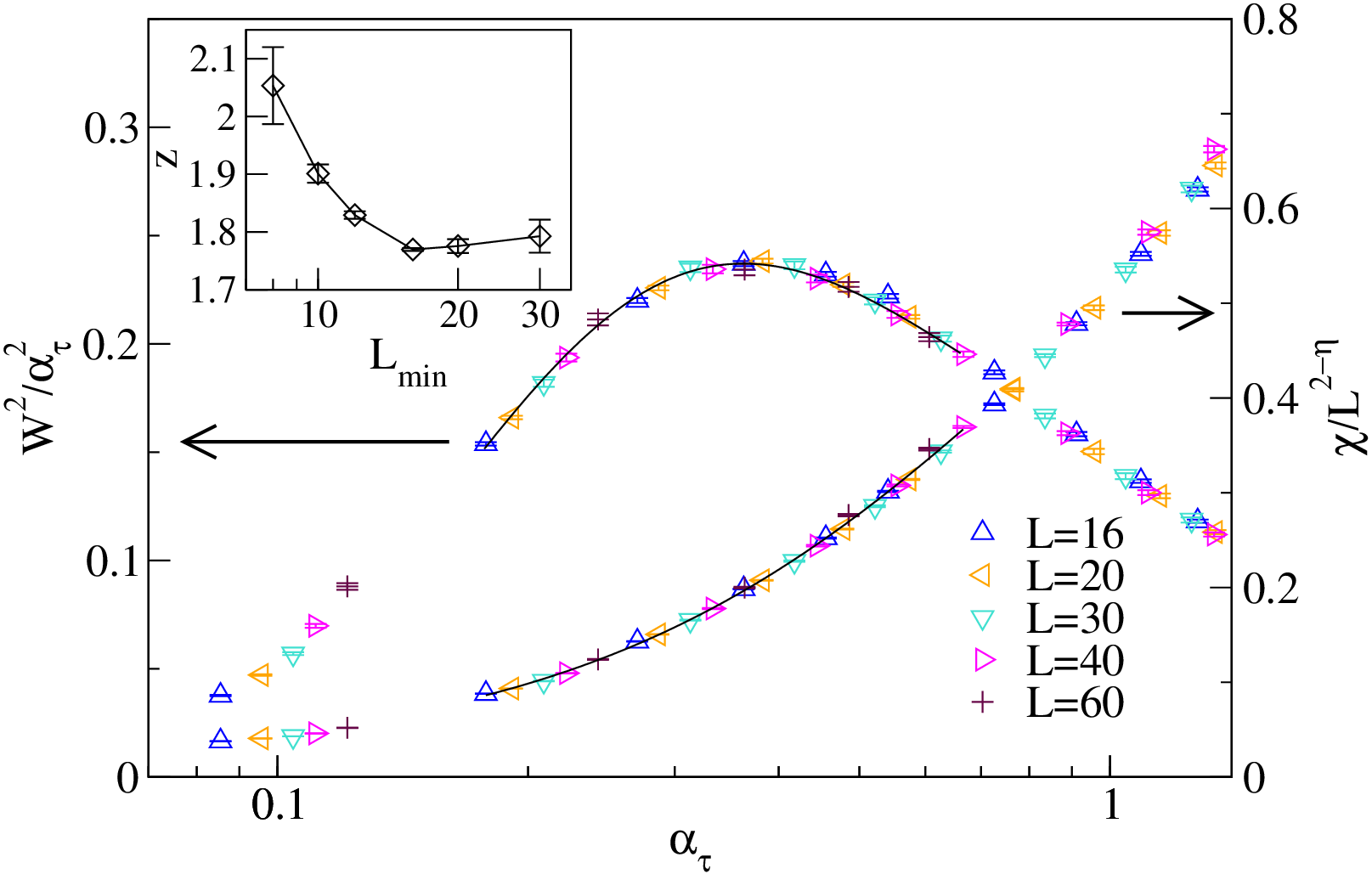}
\caption{FSS data collapses of MC data for the scaled winding number
  fluctuation $\Phi=W^2/\asp^2$ and susceptibility $\chi/L^{2-\eta}$
  as functions of $\asp=\lt/L^z$ for $\kc=0.2477, z=1.8, \eta=-0.29$.
  Solid curves are polynomial fits to the data.  Inset: Dependence of
  $z$ on the range of system sizes used in the estimate.  $L_{\rm
    min}$ indicates the smallest in a sequence of size quadruplets in
  $L=8, 10, 12, 16, 20, 30, 40, 60$ used to estimate $z$, except for
  $L_{\rm min}=30$ which indicates sizes $30, 40, 60$.}
\label{fig4}
\end{figure}

Finally we systematically study the system size dependence of the
estimated exponents and estimate errors.  This final calculation does
not involve the maxima, and thus avoids any errors in their
determination.  A double polynomial expansion is done of the scaling
functions in Eq.\ (\ref{eq:phiscaling}) in both arguments.  The
parameters are determined by $\chi^2$-minimization of the RMS
deviations of the MC data points from the scaling function.  
We performed several fits for MC data points selected from different
intervals in the range $0.2 < \asp < 1.2$ in order to verify the
stability of the results.
To study system size trends of the results, fits were made for a
sequence of system size quadruplets in $L=8, 10, 12, 16, 20, 30, 40,
60$.  The result for $z$ is shown in the inset of Fig.\ \ref{fig4}.
The displayed trend agrees with the one indicated in Fig.\ \ref{fig2}.
For the fit with $L=16,20,30,40$, $0.2 < \asp < 0.5$, we get
$\chi^2/{\rm DOF} \approx 0.8$.  Our final estimates including error
estimates based on statistical errors determined by the bootstrap
method combined with average variations from the dependence on the
$\asp$-interval included in the fits are $K_c=0.2477 \pm 0.0002, z=1.8
\pm 0.05, \nu=1.15 \pm 0.03$, and $\eta=-0.3 \pm 0.1$.  Other critical
exponents can be estimated from these values using scaling laws.

{\em Discussion -- } Analysis of our MC data of the 2D boson
localization transition by disorder revises previous estimates of the
critical exponents.  In particular the dynamic critical exponent is
estimated to $z=1.8 \pm 0.05$, which suggests that $z=d$ is not
fulfilled in $d=2$, although the values are close.  Our results
clarify how most previous simulations appear consistent with $z=2$.
For small system sizes $z=2$ works quite well, but including larger
sizes reveals corrections to scaling making $z=1.8$ a better estimate.
Our estimates are quite different from those of Ref.\
\cite{Priyadarshee:2006p740}, which we believe may be explained by
their smaller disorder averaging and uncertainty in their location of
the quantum critical point.  
The prediction of a universal conductivity
at the transition is independent of the value of $z$
\cite{Fisher:1989p868}.  However, actual estimates of the universal
value of the conductivity indirectly depend on the value of $z$, and
should be reexamined in the light of the present results.
A better analytic understanding of the
quantum critical dynamics as well as further experimental measurements
probing these issues would be welcome.

We acknowledge valuable discussions with Steve Girvin, Steve Teitel,
and Igor Herbut.  This project was supported by the Swedish Research
Council and by the Swedish National Infrastructure for Computing via
PDC.

\end{document}